\documentclass[journal]{IEEEtran}

\usepackage{xcolor,soul,framed} 

\colorlet{shadecolor}{yellow}
\usepackage[pdftex]{graphicx}
\graphicspath{{../pdf/}{../jpeg/}}
\DeclareGraphicsExtensions{.pdf,.jpeg,.png}

\usepackage[cmex10]{amsmath}
\usepackage{array}
\usepackage{mdwmath}
\usepackage{mdwtab}
\usepackage{eqparbox}
\usepackage{url}
\usepackage{graphicx}
\usepackage{cite}
\usepackage{amsmath,amssymb,amsfonts}
\usepackage{algorithm}
\usepackage{algorithmicx}
\usepackage[noend]{algpseudocode}
\usepackage{graphicx}
\usepackage{textcomp}
\usepackage{multirow}
\usepackage{subcaption}
\usepackage{makecell}
\usepackage{hyperref}

\newtheorem{defn}{Definition}[section]

\def\BibTeX{{\rm B\kern-.05em{\sc i\kern-.025em b}\kern-.08em
    T\kern-.1667em\lower.7ex\hbox{E}\kern-.125emX}}

\makeatletter
\newcommand{\ostar}{\mathbin{\mathpalette\make@circled\star}}
\newcommand{\make@circled}[2]{%
  \ooalign{$\m@th#1\smallbigcirc{#1}$\cr\hidewidth$\m@th#1#2$\hidewidth\cr}%
}
\newcommand{\smallbigcirc}[1]{%
  \vcenter{\hbox{\scalebox{0.77778}{$\m@th#1\bigcirc$}}}%
}

\newcommand{\R}{\mathbb{R}}

\newcommand{\ttilde}{{\raise.17ex\hbox{$\scriptstyle\sim$}} }

\begin{document}
\title{PocketNet: A Smaller Neural Network For Medical Image Analysis}

\author{
Adrian Celaya,
Jonas A. Actor,
Rajarajesawari Muthusivarajan,
Evan Gates,
Caroline Chung,
Dawid Schellingerhout,
Beatrice Riviere,
and David Fuentes

\thanks{Submitted for review on June 23, 2021. Jonas A. Actor and Evan Gates were both supported by a training fellowship from the Gulf Coast Consortia, on NLM Training Program in Biomedical Informatics \& Data Science (T15LM007093). David Fuentes and Dawid Schellingerhout are partially supported by R21CA249373. Beatrice Riviere is partially supported by NSF-DMS2111459. Adrian Celaya is supported by the Department of Defense through the National Defense Science \& Engineering Graduate Fellowship Program.}
\thanks{The research in this paper was supported in part by the Tumor Measurement Initiative through the MD Anderson Strategic Research Initiative Development (STRIDE) and the National Science Foundation awards NSF-2111147 and NSF-2111459.}
\thanks{Adrian Celaya is with Rice University, Houston, TX 77005 USA and the University of Texas MD Anderson Cancer Center, Houston, TX 77030 (email:aecelaya@rice.edu).}
\thanks{Rajarajesawari Muthusivarajan, Caroline Chung, and Dawid Schellingerhout are with the University of Texas MD Anderson Cancer Center, Houston, TX 77030 USA.}
\thanks{David Fuentes is with the University of Texas MD Anderson Cancer Center, Houston, TX 77030 USA and Rice University, Houston, TX 77005 USA (email: dtfuentes@mdanderson.org)}
\thanks{Jonas A. Actor was with Rice University, Houston, TX 77005 USA and the University of Texas MD Anderson Cancer Center, Houston, TX 77030 USA. He is now with Sandia National Laboratories, Albuquerque, NM 87123 USA.}
\thanks{Evan Gates was with the the University of Texas MD Anderson Cancer Center, Houston, TX 77003. He is now with the University of Washington, Seattle, WA, 98195.}
\thanks{Beatrice Riviere is with Rice University, Houston, TX 77005 USA.}
\thanks{Adrian Celaya and Jonas A. Actor contributed equally.}
}

\maketitle 
\begin{abstract}
Medical imaging deep learning models are often large and complex, requiring specialized hardware to train and evaluate these models. To address such issues, we propose the PocketNet paradigm to reduce the size of deep learning models by throttling the growth of the number of channels in convolutional neural networks. We demonstrate that, for a range of segmentation and classification tasks, PocketNet architectures produce results comparable to that of conventional neural networks while reducing the number of parameters by multiple orders of magnitude, using up to 90\% less GPU memory, and speeding up training times by up to 40\%, thereby allowing such models to be trained and deployed in resource-constrained settings.
\end{abstract}


\section{Introduction}

Deep learning is an increasingly common framework for automating and standardizing essential tasks related to medical image analysis that would otherwise be subject to wide variability. For example, delineating regions of interest (i.e., image segmentation) is necessary for computer-assisted diagnosis, intervention, and therapy \cite{intro-2}. Manual image segmentation is a tedious, time-consuming task whose results are often subject to wide variability among users \cite{intro-1, intro-3}. On the other hand, fully automated segmentation can substantially reduce the time required for target volume delineation and produce more consistent segmentation masks \cite{intro-1, intro-3}. Over the last several years, deep learning methods have produced impressive results for segmentation tasks such as labeling tumors or various anatomical structures \cite{unet, unet-3d, densenet, resnet}. 

However, the performance of deep learning methods comes at an enormous computational and monetary cost, independent of concerns about data quality. Training networks to convergence can take several days or weeks using specialized computing equipment with sufficient computing capacity and available memory to handle large imaging datasets. The cost of a workstation with suitable hardware specifications for training large deep learning models ranges from roughly \$5,700 to \$49,000, whereas a dedicated deep learning-enabled server blade can range from \$31,500 to \$134,000 \cite{lambda}. Cloud-based solutions offer a more economical option for training models by allowing users to pay for time on accelerated computing instances. However, the cost of such instances ranges from \$3 to \$32 per hour, and additional measures must be implemented to protect patient privacy \cite{aws}. The latter may involve an institution entering into a service agreement with a cloud-computing resource provider. The cost of such an agreement is another consideration that must be taken into account.  

We wish to reduce the costs - in model size, training time, and memory requirements - associated with training deep learning models while preserving their performance. To do so, we propose the PocketNet paradigm for deep learning models, \emph{a straightforward modification to existing architectures that substantially reduces the number of parameters and maintains the same performance as the original architecture}. This modification questions the long-held assumption that doubling the number of features after each downsampling operation (i.e., pooling or convolution) is necessary for convolutional neural networks \cite{lenet, imagenet, alexnet, resnet, densenet, unet-3d, unet, hrnet, nnunet, unet-common}. Our work demonstrates the effectiveness of our PocketNets in several 3D segmentation tasks and a classification problem. We also show that the PocketNet paradigm carries several practical benefits - less GPU memory required for training, less time required for training, and faster inference times. 


\subsection{Previous work}

The last several years have seen several attempts to decrease the number of parameters in deep learning architectures in medical imaging. Broadly, these attempts fall into two categories: post-processing tools and architecture design strategies. More specifically, pruning (post-processing tool), depth-wise separable (DS) convolutions (architecture design strategy), and filter reductions (architecture design strategy) are known to help mitigate the overparameterization of deep convolutional neural networks (CNNs) for 3D medical image segmentation \cite{pruning1, ds-conv-primer, filter-reduce1}. These methods, alone and combined, give rise to many of the novel, efficient deep learning architectures that are currently popular. We briefly survey these network reduction strategies below.

Introduced by LeCun et al., the purpose of pruning is to remove the redundant connections within a neural network \cite{pruning1}. Pruning starts with a large pre-trained model and involves deleting weights and iteratively retraining the model until a significant drop in performance occurs. Pruning in medical image analysis reduces the required inference time and GPU memory while maintaining model performance \cite{unet++-pruning, NASUnet-pruning, medseg-pruning}. However, network pruning is a post-processing step that is applied to an existing pre-trained network; although useful, pruning does not solve the demands of training large, overparameterized models, which requires a great deal of memory and high-end computing hardware.

Network architecture design strategies for reducing the number of parameters in deep neural networks for medical image analysis include DS convolutions and reduction of the number of feature maps at each layer. DS convolutions are well known for reducing the number of parameters associated with convolution \cite{ds-conv-primer, xception, mobilenet}. DS convolution factorizes the standard convolution operation into two distinct steps: depth-wise convolution followed by a point-wise convolution. A normal convolution layer has a number of parameters that is quadratic in the number of channels. In contrast, DS convolutions have a number of parameters that is linear in the number of channels. In practice, recently developed medical neural network architectures take advantage of DS convolution to reduce the number of parameters in their networks by up to a factor of 5 while achieving results comparable with those of conventional convolution \cite{ds-conv-1, ds-conv-2}. However, 3D DS convolution is not supported in standard deep learning packages and requires more memory than standard convolution during training due to the storage of an extra gradient layer.

By convention, the number of feature maps doubles after each downsampling operation in a CNN. This growth in the number of feature maps accounts for a large portion of the number of training parameters. With this in mind, perhaps the simplest way to reduce overparameterization in a deep learning model is to reduce the number of feature maps. Van der Putten et al. explore this idea in \cite{filter-reduce1} by dividing the number of feature maps in the decoder portion of a U-Net by a constant factor $r$. For increasing values of $r$, the number of parameters in the decoder is reduced by up to a factor of 100, and segmentation performance remains the same. However, this approach introduces another hyperparameter $r$ that needs to be tuned, does not reduce the number of parameters in the encoder branch of the U-Net architecture, and is only applicable to segmentation models.

\subsection{Novel Contributions}
This paper makes the following novel contributions:
\begin{enumerate}
    \item We propose the PocketNet paradigm. This paradigm builds from the definition of multigrid methods from numerical linear algebra. (Section \ref{sec:pocketdef})
    \item We demonstrate that PocketNet architectures perform comparably with their conventional neural network architecture counterparts. We measure PocketNet performance on three segmentation problems and one classification problem, using data from recent public medical imaging challenges. (Section \ref{sec:acc})
    \item We profile the memory footprint for training conventional and PocketNet architectures, highlighting that PocketNet reduces the memory footprint for training models. During this profiling, we also track training time per epoch, showing that PocketNet models take less wall-time to train than do their conventional counterparts. (Section \ref{sec:gpu-usage})
    \item We perform a controlled study of the effects of doubling the number of feature maps in convolutional neural networks, and additionally compare the intensities of the learned activations for voxel-wise classification (Sections \ref{sec:ablation} and \ref{sec:feature-activation}). To our knowledge, this constitutes the first controlled study of the effects of feature map doubling in convolutional deep learning architectures for medical imaging tasks.
\end{enumerate}

\section{Materials and Methods}
\label{sec:methods}

\subsection{The PocketNet Paradigm}
We propose a modification to existing network architectures that dramatically reduces the number of parameters while also retaining performance. Many common network architectures for imaging tasks rely on manipulating images (or image features) at multiple scales because natural images encapsulate data at multiple resolutions. As a result, most CNN architectures -- including many popular state-of-the-art methods such as nnUNet \cite{nnunet} and HRNet \cite{hrnet} --  follow a pattern of downsampling and upsampling, following the intuition of the original U-Net paper \cite{unet} that popularized this approach. In the architecture first presented there, the number of feature maps (i.e., channels) in each convolution operator is doubled each time the resolution of the images decreases; the justification being that the increased number of feature maps offsets the loss of information from downsampling. This idea, of compensating for lost information by increasing the number of features, can be traced before U-Net, to the original ImageNet paper \cite{imagenet} and earlier. In all of these architectures, from ImageNet to U-Net to the state-of-the-art architectures today, the recurring refrain is that training is limited by compute availability due to the network's size, since the number of parameters in all of these architectures grows exponentially as the number of channels is doubled.

However, other classical methods that manipulate images (or other signals) at multiple scales assume a hierarchy of scales i.e. that information can be decomposed into coarse-scale and fine-scale features \emph{independently}; most prominent among such methods are those based on wavelets \cite{siambook} and on multigrid methods \cite{multigrid}. 
Intrinsic to these methods is the construction of a series of grids of appropriate resolution. At each resolution, the constructed grid allows for the specific frequencies to be resolved, without aliasing.
Because of this hierarchy of scales, and that specific sets of frequencies are resolved by specific grids at specific scales, the information capacity of coarser scales is guaranteed to be less than that of finer scales, and as a result, fewer operations (for multigrid solvers) and memory (for wavelets) are required: information ``lost'' by downsampling into a smaller, coarser subspace is accounted for at a grid of different resolution, and when images are downsampled to a coarser resolution, it is not necessary to double the number of channels or dimensions to preserve the information capacity at each downsampling instance. 
As a result, we propose that the doubling of the channels at each resolution in CNN architectures like U-Net is not intrinsically necessary, since each depth in these architectures corresponds to features at a different scale.

A generic U-Net framework is fully written out in Algorithm \ref{fig:Unet}.
\begin{figure}[t!]
\begin{algorithm}[H]
\caption{\label{fig:Unet}
U-Net \\
 \textit{Input}: Tensor $u_D$, integers $D, \nu$}
\begin{algorithmic}
\Procedure{Block}{$\nu, u$} \Comment{Tensor $u$, integer $\nu$}
\For{$i = 1:\nu$}
    \State $u \leftarrow \sigma {\color{white}~^{-1}} \left( K_i * u \,\, + b_i \right)$ \Comment{Convolution block}
\EndFor
\State \Return $u$
\State
\EndProcedure
\Procedure{U-Net}{$u_D, D, \nu$}
\State $u_D \leftarrow $ \texttt{Block}$(\nu, u_D)$ \Comment{Initial computation}
\State
\State
\State \emph{Encoder}
\For{$d = D - 1$ \text{to} $1$}
    \State $u_d \leftarrow \Pi_{d + 1}^{d} u_{d + 1}$ \Comment{Downsample}
    \State $u_{d} \leftarrow$ \texttt{Block}$(\nu, u_d)$
    \State {\color{white} $r_d \rightarrow f_d - A_d$}
\EndFor
\State
\State \emph{Coarsest resolution}
\State $v_{0} \leftarrow \Pi_{1}^{0}u_{1}$ 
\State $v_{0} \leftarrow$ \texttt{Block}$(\nu, v_{0})$ 
\State
\State \emph{Decoder}
\For{$d = 1$ \text{to} $D$}
    \State $v_d \leftarrow \Pi_{d-1}^d v_{d-1}$ \Comment{Upsample}
    \State $v_d \leftarrow u_d + v_d$ \Comment{Skip connection}
    \State $v_{d} \leftarrow$ \texttt{Block}$(\nu, v_d)$
\EndFor
\State \Return $\phi(K_{\text{out}} * v_D + b_{\text{out}})$
\EndProcedure
\end{algorithmic}
\end{algorithm}
\end{figure}
In the \texttt{Block} procedure of this algorithm, each convolution kernel $K_i$ has some number of channels-in and channels-out that depend on the layer's depth in the `U' of the architecture. The overall depth $D$ used in this architecture is commonly set to $D = 3$ to $D = 6$ \cite{unet, unet-3d}. If the convolutions have $c_{\text{in}}$ channels-in and $c_{\text{out}}$ channels-out at the network's finest resolution, the convolutions in the next layer will have $2 c_{\text{in}}$ channels-in and $2 c_{\text{out}}$ channels-out. Subsequently, at resolution depth $d$, there are $2^d c_{\text{in}}$ channels-in and $2^d c_{\text{out}}$ channels-out for the convolutions. As a result, the number of parameters in a CNN grows exponentially with increasing depth, and this exponential growth is the driving factor for the large size of image segmentation networks.

We remark that Algorithm \ref{fig:Unet} is nearly identical to a single V-cycle of a geometric multigrid solver for solving the linear system of equations $Au = f$, where the linear system $A$ and the unknown variable $u$ relate to a geometric grid involving multiple resolutions \cite{multigrid, saad2003iterative}.
An algorithm for a V-cycle is shown in Algorithm \ref{fig:vcycle}.
\begin{figure}[t]
\begin{algorithm}[H]
    \caption{\label{fig:vcycle}Multigrid V-Cycle, adapted from \cite{saad2003iterative} \\
    \textit{Input}: Matrix $A_D$, vectors $f_D, u_D$, integers $D, \nu$}
\begin{algorithmic}
\Procedure{Block}{$\nu, A, u, f$}           \Comment{Matrix $A$, vectors $u, f$}
\For{$i = 1:\nu$}                           
    \State $u \leftarrow D^{-1}(f - (A-D)u)$ \Comment{Iterative step}
\EndFor
\State \Return $u$
\State
\EndProcedure
\Procedure{V-Cycle}{$A_D, u_D, f_D, D, \nu$}
\State $u_D \leftarrow $ \texttt{Block}$(\nu, A_D, u_D, f_D)$ \Comment{Initial computation}
\State $r_D \leftarrow f_D - A_D u_D$
\Comment{Residual update}
\State
\State \emph{Encoder}
\For{$d = D - 1$ \text{to} $1$}
    \State $f_d, A_d \leftarrow \Pi_{d + 1}^{d} r_{d + 1}, A_{d + 1}$ \Comment{Downsample}
    \State $u_d \leftarrow $ \texttt{Smooth}$(\nu, A_d, 0, f_d)$
    \State $r_d \leftarrow f_d - A_d u_d$
    \Comment{Residual update}
\EndFor
\State
\State \emph{Coarsest resolution}
\State $f_0, A_0 = \Pi_{1}^{0}r_1, A_1$ 
\State $v_{0} = A_{0}^{-1} f_0$ \Comment{Direct solve}
\State
\State \emph{Decoder}
\For{$d = 1$ \text{to} $D$}
    \State $v_d \leftarrow \Pi_{d-1}^d v_{d-1}$ \Comment{Upsample}
    \State $v_d \leftarrow u_d + v_d$ \Comment{Skip connection}
    \State $v_d \leftarrow $ \texttt{Block}$(\nu, A_d, u_d, f_d)$
\EndFor
\State \Return $v_D$ {\color{white} $ \left( K_{\text{out}} \right) $ }
\EndProcedure
\end{algorithmic}
\end{algorithm}
\end{figure}

\begin{figure*}[ht!]
    \centering
    \includegraphics[width=\textwidth]{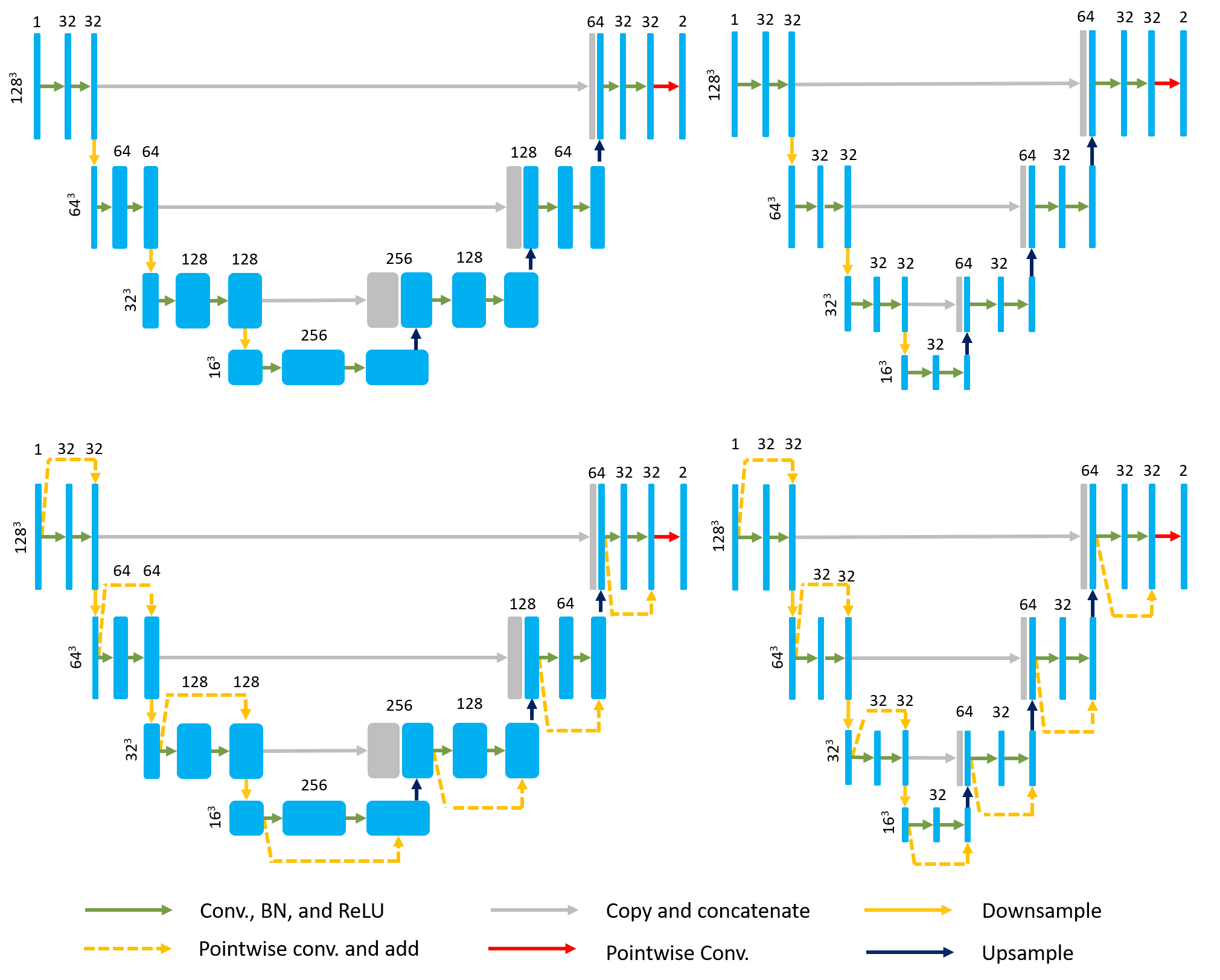}
    \caption{(Left) Full U-Net (top) and ResNet (bottom) segmentation architectures where for each downsampling step, the number of feature maps doubles after each convolution layer. (Right) Pocket U-Net (top) and ResNet (bottom) segmentation architecture. In this PocketNet architecture, the number of feature maps resulting from each convolution layer remains constant regardless of spatial resolution, resulting in substantially fewer parameters overall.\label{fig:full_pocket}}
\end{figure*}

The PocketNet paradigm, defined in Definition \ref{def:pocketnet}, exploits this similarity between multigrid methods and U-Net-like architectures. This paradigm proposes that the number of feature maps used at the finest resolution is sufficiently rich to capture the relevant information for the imaging task at hand and that doubling the number of channels is unnecessary.
Instead of doubling the number of feature maps at every level of a CNN, we keep them constant, substantially reducing the number of parameters in our models in the process. We designate network architectures that keep the number of feature maps constant as \textit{Pocket Networks}, or \textit{PocketNets} for short, in the sense that these networks are ``small enough to fit in one's back pocket''. For example, ``Pocket U-Net'' refers to a U-Net architecture where we apply our proposed modification. Figure \ref{fig:full_pocket} provides a visual representation of the PocketNet paradigm applied to a U-Net. 
We present a formalized definition of the PocketNet paradigm in Definition \ref{def:pocketnet}.
~\\
\begin{defn}\label{def:pocketnet}
A network architecture obeys the \textit{PocketNet paradigm} if the range of all convolution operators (except the final output layer) present in the network is a subset of $\R^{h \times w \times c_{\text{out}}}$, where $c_{\text{out}}$ is fixed. Such a network is called a Pocket Network, or PocketNet for short.
\end{defn} ~\\

While the relationship between multigrid algorithms and U-Nets has been explored (see \cite{mgnet} and references therein), the PocketNet paradigm is applicable to any CNN architecture, regardless of whether or not the architecture is similar in overall structure to a U-Net.

The number of parameters saved in the PocketNet architectures is substantial. Assume that a PocketNet and its corresponding full-sized architecture are identical apart from the doubling of the number of channels in standard convolutions. For simplicity's sake, we further assume that the number of convolutions performed at each resolution is the same, denoted by $C$. Also for simplicity's sake, we assume that the window of each convolution kernel is isotropic, with a stencil width of $k$ in each dimension. We denote the maximum resolution depth, i.e. the number of downsampling operations in the network architecture, as $D$. In a full-sized network, a convolution kernel at resolution depth $d$ operating on an $n$D image (for $n=2$ or $n=3$) with a stencil width $k$ has $k^n \left( c_{\text{in}} 2^d \right) \left( c_{\text{out}} 2^d \right)$ parameters. Therefore, the full-sized network has the following number of parameters 
\begin{equation} n_{\text{full}} = \sum_{d=0}^D C k^n (c_{\text{in}}2^d)(c_{\text{out}}2^d) = \frac{1}{3} C k^n c_\text{in} c_{\text{out}} \left( 4^{D+1} - 1 \right).
\end{equation}
On the other hand, the convolution kernels in a PocketNet architecture have $k^n c_{\text{in}} c_{\text{out}}$ parameters at each resolution depth, resulting in the following number of parameters
\begin{equation}
n_\text{pocket} = \sum_{d=0}^D C k^n c_\text{in} c_\text{out} = (D+1)Ck^n c_\text{in} c_\text{out}.
\end{equation}
Therefore, the PocketNet paradigm reduces the number of parameters in a network by up to a factor of 
\begin{equation}
\text{savings} = \frac{n_{\text{full}}}{n_{\text{pocket}}} 
= \frac{4^{D+1} - 1}{3 (D+1)}
\approx \frac{4^D}{D}.
\end{equation}
This analysis highlights that the growth of parameters with increasing depth is exponential for full networks but only linear for PocketNets.

\label{sec:pocketdef}

\subsection{Experiments}

\subsubsection{Data}
\label{sec:data}
We test PocketNet on a range of recent public medical imaging challenge datasets. These datasets encompass three segmentation problems and one classification task, all with various data set sizes, complexity, and modalities. Two of our segmentation tasks - binary liver segmentation in the MICCAI Liver and Tumor Segmentation (LiTS) Challenge 2017 dataset \cite{lits} and single-contrast brain extraction in the Neurofeedback Skull-stripped (NFBS) repository \cite{nfbs} - are comparatively simple. We use these datasets for baseline comparisons, much like the MNIST Fashion and CIFAR-10 datasets are used to evaluate newly proposed classification architectures. The third segmentation task - multilabel tumor segmentation in the MICCAI Brain Tumor Segmentation (BraTS) Challenge 2020 dataset \cite{brats1, brats2, brats3} - is commonly viewed as a more complex and technical segmentation problem than LiTS or NFBS segmentation. Therefore, we use BraTS data for our performance benchmarks (see Section \ref{sec:gpu-usage}). Finally, our classification task - binary classification in the COVIDx8B dataset \cite{covidx} - evaluates the PocketNet paradigm with a much larger dataset, demonstrating its appropriateness for pertinent problems other than image segmentation. These datasets and their related pre-processing and post-processing methods are described below.

\paragraph{LiTS Data}
For the LiTS dataset, we perform binary liver segmentation. This dataset consists of the 131 CT scans from the MICCAI 2017 Challenge's multi-institutional training set. These scans vary significantly in the number of slices in the axial direction and voxel resolution, although all axial slices are at 512$\times$512 resolution. As a result, we use the preprocessing steps proposed by nnUNet to handle this variability \cite{nnunet}. We resample each image to the median resolution of the training data in the $x$ and $y$-directions and use the 90th percentile resolution in the $z$-direction. For intensity normalization, we window each image according to the foreground voxels' 0.5 and 99.5 percentile intensity values across all of the training data. This scheme results in windowing from -17 to 201 HU. We also apply z-score normalization according to the foreground voxels' mean and standard deviation. The LiTS dataset is available for download \url{https://competitions.codalab.org/competitions/17094#learn_the_details-overview}.

\paragraph{NFBS Data} 
The segmentation task for the NFBS dataset is extraction (i.e., segmentation) of the brain from MR data. The NFBS dataset consists of 125 T1-weighted MR images with manually labeled ground truth masks. All images are provided with an isotropic voxel resolution of 1$\times$1$\times$1 $\text{mm}^3$ and are of 256$\times$256$\times$192 resolution. For pre-processing, we apply z-score intensity normalization. The NFBS dataset is available for download at \url{http://preprocessed-connectomes-project.org/NFB_skullstripped/}.

\paragraph{BraTS Data}
The BraTS training set contains 369 multimodal scans from 19 institutions. Each set of scans includes a T1-weighted, post-contrast T1-weighted, T2-weighted, and T2 Fluid Attenuated Inversion Recovery volume and a multilabel ground truth segmentation. We merge the labels in each ground truth segmentation and perform whole tumor segmentation for our analysis. All volumes are provided at an isotropic voxel resolution of 1$\times$1$\times$1 $\text{mm}^3$, co-registered to one another, and skull stripped, with a size of 240$\times$240$\times$155. We crop each image according to the brainmask (i.e., non-zero voxels) and apply z-score intensity normalization on only non-zero voxels for pre-processing. The BraTS training dataset is available for download at \url{https://www.med.upenn.edu/cbica/brats2020/registration.html}. 

\paragraph{COVIDx8B Data} The classification task for the COVIDx8B dataset is COVID-19 detection on 2D chest x-rays. The COVIDx8B dataset consists of a training set with 15,952 images and an independent test set with 400 images. The training set is class-imbalanced, with  13,794 COVID-19 negative images and 2,158 COVID-19 positive images. However, the COVIDx8B test set is class-balanced, with 200 COVID-19–negative and 200 COVID-19–positive images. We resize each image to a resolution of 256$\times$256 and apply z-score intensity normalization as pre-processing steps. The COVIDx8B dataset is available for download at \url{https://github.com/lindawangg/COVID-Net/blob/master/docs/COVIDx.md}.

\subsubsection{Network Architectures}
\label{sec:architectures}
For each of our tasks and datasets, we compare the performance of various full-sized architectures with their PocketNet counterparts. 



\begin{figure}[ht!]
    \centering
    \includegraphics[width=0.9\columnwidth]{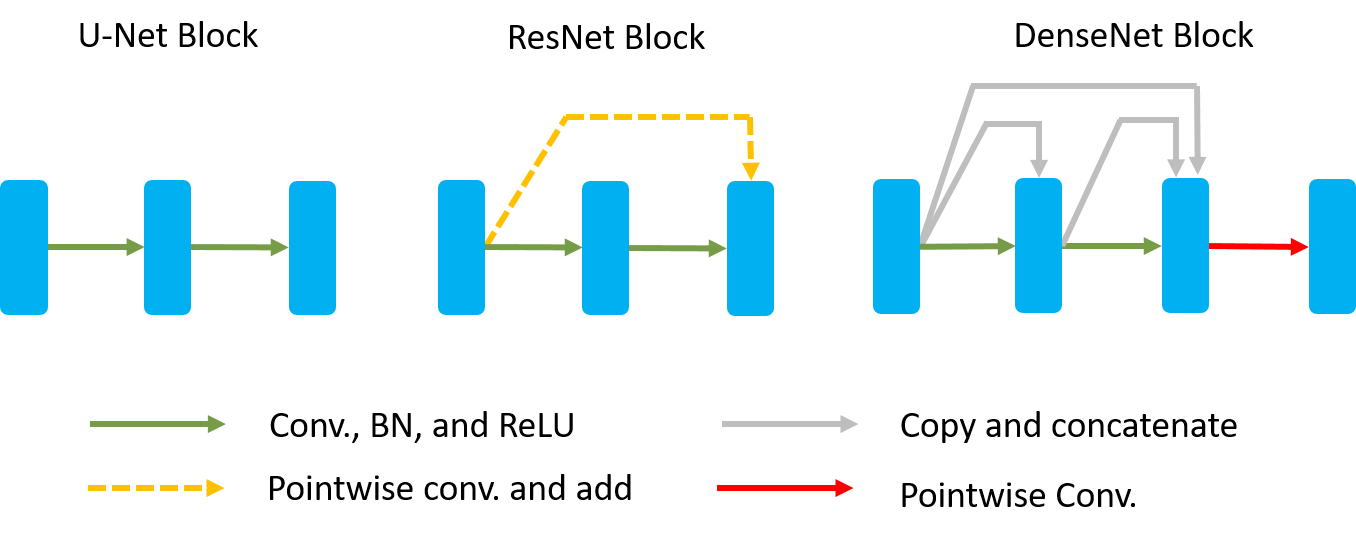}
    \caption{Block designs for U-Net (left), ResNet (center), and DenseNet (right).}
    \label{fig:blocks}
\end{figure}

\paragraph{Segmentation Architectures}
We examine the effects of our proposed modification strategy using five segmentation architectures - U-Net \cite{unet-3d}, ResNet \cite{resnet}, DenseNet \cite{densenet}, HRNet \cite{hrnet}, and nnUNet \cite{nnunet}. The U-Net, ResNet, and DenseNet architectures are similar. They use the same U-Net backbone with variations in their convolution blocks (see Figure \ref{fig:blocks}) but are highly prevalent architectures in the literature for 3D medical image segmentation \cite{unet-common}. HRNet is not as prevalent in the literature as U-Net and its variants but does represent a fundamentally different deep learning architecture for 3D segmentation. Rather than using a "U" shape, where we continually coarsen (i.e., downsample via pooling) high-resolution features, HRNet preserves features at each resolution level. The nnUNet architecture provides a state-of-the-art baseline for our analysis.


\paragraph{COVIDx8B Architectures}
Our classification architecture for the COVIDx8B dataset is a U-Net encoder with four downsampling layers. The final layer of the U-Net encoder is flattened via global max-pooling and passed to a fully connected layer for the network's final output. Visually, this architecture is represented by the left half of the Pocket U-Net architecture in Figure \ref{fig:full_pocket}.

\subsubsection{Training Protocols and Hyperparameters}
\label{sec:training}
For each task (e.g., segmentation and classification), we initialize the first layer in each network with 32 feature maps and use the Adam optimizer \cite{adam}. The learning rate is set to 0.0003. Training for all segmentation tasks uses a batch size of two and a batch size of 32 for COVIDx8B classification. For all segmentation tasks, we use a patch size of 128 $\times$ 128 $\times$ 128 and apply the same random augmentation described in \cite{nnunet}. To evaluate a predicted segmentation mask's validity, we use the Sorensen-Dice Similarity Coefficient (Dice), the 95$^{\text{th}}$ percentile Hausdorff distance, and the average surface distance. Implementations of these metrics are available through the SimpleITK Python package \cite{simpleitk1, simpleitk2, simpleitk3}. For the segmentation tasks, our loss function is calculated as an $L_2$ relaxation of the Dice score; for a true segmentation $Y_{true}$ and a predicted segmentation $Y_{pred}$, our $L_2$-Dice loss function is taken from \cite{actor2020identification} and is given as 
\begin{equation}
Loss_{Dice}(Y_{true}, Y_{pred}) = \frac{\left \Vert Y_{true} - Y_{pred} \right \Vert_2^2}{\left \Vert Y_{true} \right \Vert_2^2 + \left \Vert Y_{pred} \right \Vert_2^2}.
\end{equation} 
For the nnUNet architecture, we add binary cross entropy to the Dice loss shown above.

For the classification tasks, we use categorical cross-entropy as our loss function, with outputs being of two classes (COVID-19 positive and negative). We use the receiver operating characteristic area under the curve (AUC) metric to evaluate each classification model's validity. This metric is available via the scikit-learn Python package. Our models are implemented in Python using TensorFlow (v2.8.0) and trained on an NVIDIA Quadro RTX 8000 GPU \cite{keras}. All network weights are initialized using the default initializers from TensorFlow. All other hyperparameters are left at their default values. The code for each network architecture is available at \url{github.com/aecelaya/pocketnet}.

\subsubsection{Inference and Post-Processing}
\label{sec:inference}
We perform inference on test images using a sliding window approach for each segmentation task, where the window size equals the patch size set during training. After each window prediction, we slide the window by half the size of the patch. Additionally, we apply a Gaussian importance weighting ($\sigma = 0.125$) to each window prediction \cite{nnunet}. 

For post-processing, we try each of the following strategies for each segmentation task and network:
\begin{enumerate}
    \item Apply morphological clean-up - i.e., erode by two voxels, retain the largest connected component, dilate by one voxel, and fill holes.
    \item Only retains the largest connected component.
    \item Do nothing.
\end{enumerate}
We select the strategy that yields the highest mean Dice as our final post-processing strategy for a given task and network.

We note that the inference and post-processing steps described above are applied to \emph{all} segmentation tasks for every network we test.

\section{Results}
\label{sec:results}


\subsection{Comparable Accuracy}
\label{sec:acc}
We use the training parameters described in Section \ref{sec:training} to train the architectures listed in Section \ref{sec:architectures}. For the segmentation tasks described in Section \ref{sec:data}, we employ a five-fold cross-validation scheme to obtain predictions for each dataset and architecture. For classification on the COVIDx8B dataset, we train each model with the training set and generate predictions on the test set. These results of these experiments are shown in Tables \ref{fig:accuracy-table-seg} and \ref{tab:accuracy-table-class}.

\begin{table*}[ht!]
\centering
\resizebox{\textwidth}{!}{%
\begin{tabular}{lllcccc|ccc|ccc}
\hline
\multicolumn{1}{c}{\multirow{2}{*}{Task}} &
  \multicolumn{1}{c}{\multirow{2}{*}{Architecture}} &
  \multicolumn{1}{c}{\multirow{2}{*}{Variant}} &
  \multirow{2}{*}{\begin{tabular}[c]{@{}c@{}}\# Params \\ (M)\end{tabular}} &
  \multicolumn{3}{c|}{Dice} &
  \multicolumn{3}{c|}{Hausdorff 95 (mm)} &
  \multicolumn{3}{c}{Avg. Surface (mm)} \\ \cline{5-13} 
\multicolumn{1}{c}{} &
  \multicolumn{1}{c}{} &
  \multicolumn{1}{c}{} &
   &
  Mean (Std) &
  Median &
  p-value &
  Mean (Std) &
  Median &
  p-value &
  Mean (Std) &
  Median &
  p-value \\ \hline
\multirow{10}{*}{LiTS} &
  \multirow{2}{*}{U-Net} &
  Full &
  90.5 &
  0.922 (0.098) &
  0.952 &
  \multirow{2}{*}{NS} &
  9.715 (23.95) &
  0.967 &
  \multirow{2}{*}{NS} &
  2.713 (3.637) &
  1.446 &
  \multirow{2}{*}{NS} \\
 &
   &
  Pocket &
  0.8 &
  0.930 (0.059) &
  0.952 &
   &
  7.845 (21.67) &
  0.0959 &
   &
  2.343 (2.832) &
  1.384 &
   \\ \cline{2-13} 
 &
  \multirow{2}{*}{ResNet} &
  Full &
  91.9 &
  0.932 (0.097) &
  0.958 &
  \multirow{2}{*}{NS} &
  9.105 (28.40) &
  0.729 &
  \multirow{2}{*}{NS} &
  2.528 (4.876) &
  1.185 &
  \multirow{2}{*}{NS} \\
 &
   &
  Pocket &
  0.8 &
  0.928 (0.104) &
  0.956 &
   &
  10.38 (26.90) &
  0.742 &
   &
  2.615 (3.915) &
  1.262 &
   \\ \cline{2-13} 
 &
  \multirow{2}{*}{DenseNet} &
  Full &
  135.9 &
  0.933 (0.056) &
  0.955 &
  \multirow{2}{*}{NS} &
  18.97 (50.65) &
  0.781 &
  \multirow{2}{*}{NS} &
  3.478 (7.297) &
  1.234 &
  \multirow{2}{*}{*} \\
 &
   &
  Pocket &
  1.3 &
  0.930 (0.078) &
  0.951 &
   &
  10.16 (27.21) &
  1.000 &
   &
  2.588 (3.463) &
  1.405 &
   \\ \cline{2-13} 
 &
  \multirow{2}{*}{HRNet} &
  Full &
  6.5 &
  0.939 (0.067) &
  0.958 &
  \multirow{2}{*}{*} &
  6.126 (19.37) &
  0.756 &
  \multirow{2}{*}{NS} &
  2.082 (2.889) &
  1.202 &
  \multirow{2}{*}{NS} \\
 &
   &
  Pocket &
  0.7 &
  0.937 (0.055) &
  0.954 &
   &
  8.805 (26.97) &
  0.787 &
   &
  2.455 (3.931) &
  1.304 &
   \\ \cline{2-13} 
 &
  \multirow{2}{*}{nnUNet} &
  Full &
  31.2 &
  0.933 (0.092) &
  0.953 &
  \multirow{2}{*}{*} &
  6.187 (19.09) &
  0.800 &
  \multirow{2}{*}{NS} &
  2.300 (4.030) &
  1.339 &
  \multirow{2}{*}{*} \\
 &
   &
  Pocket &
  0.8 &
  0.944 (0.039) &
  0.956 &
   &
  2.844 (6.285) &
  0.785 &
   &
  1.738 (1.432) &
  1.263 &
   \\ \hline
\multirow{10}{*}{BraTS} &
  \multirow{2}{*}{U-Net} &
  Full &
  90.5 &
  0.889 (0.090) &
  0.915 &
  \multirow{2}{*}{NS} &
  5.627 (13.59) &
  1.414 &
  \multirow{2}{*}{NS} &
  1.695 (2.242) &
  1.081 &
  \multirow{2}{*}{*} \\
 &
   &
  Pocket &
  0.8 &
  0.891 (0.088) &
  0.914 &
   &
  5.089 (12.18) &
  1.414 &
   &
  1.750 (3.375) &
  1.023 &
   \\ \cline{2-13} 
 &
  \multirow{2}{*}{ResNet} &
  Full &
  91.9 &
  0.902 (0.083) &
  0.927 &
  \multirow{2}{*}{NS} &
  4.648 (11.59) &
  1.000 &
  \multirow{2}{*}{NS} &
  1.469 (2.012) &
  0.842 &
  \multirow{2}{*}{NS} \\
 &
   &
  Pocket &
  0.8 &
  0.904 (0.077) &
  0.927 &
   &
  4.007 (10.34) &
  1.000 &
   &
  1.308 (1.543) &
  0.834 &
   \\ \cline{2-13} 
 &
  \multirow{2}{*}{DenseNet} &
  Full &
  135.9 &
  0.887 (0.106) &
  0.917 &
  \multirow{2}{*}{*} &
  5.833 (12.97) &
  1.414 &
  \multirow{2}{*}{NS} &
  1.693 (2.560) &
  0.938 &
  \multirow{2}{*}{NS} \\
 &
   &
  Pocket &
  1.3 &
  0.889 (0.103) &
  0.917 &
   &
  5.935 (12.98) &
  1.414 &
   &
  1.730 (2.792) &
  0.929 &
   \\ \cline{2-13} 
 &
  \multirow{2}{*}{HRNet} &
  Full &
  6.5 &
  0.900 (0.077) &
  0.926 &
  \multirow{2}{*}{*} &
  5.729 (13.95) &
  1.000 &
  \multirow{2}{*}{NS} &
  1.659 (2.315) &
  0.927 &
  \multirow{2}{*}{NS} \\
 &
   &
  Pocket &
  0.7 &
  0.897 (0.083) &
  0.922 &
   &
  5.202 (12.37) &
  1.000 &
   &
  1.523 (2.042) &
  0.900 &
   \\ \cline{2-13} 
 &
  \multirow{2}{*}{nnUNet} &
  Full &
  31.2 &
  0.907 (0.076) &
  0.929 &
  \multirow{2}{*}{NS} &
  3.472 (8.733) &
  1.000 &
  \multirow{2}{*}{NS} &
  1.150 (1.484) &
  0.750 &
  \multirow{2}{*}{NS} \\
 &
   &
  Pocket &
  0.8 &
  0.908 (0.070) &
  0.931 &
   &
  3.586 (10.57) &
  1.000 &
   &
  1.194 (2.241) &
  0.749 &
   \\ \hline
\multirow{10}{*}{NFBS} &
  \multirow{2}{*}{U-Net} &
  Full &
  90.5 &
  0.988 (0.002) &
  0.988 &
  \multirow{2}{*}{**} &
  0.000 (0.000) &
  0.000 &
  \multirow{2}{*}{NS} &
  0.419 (0.124) &
  0.384 &
  \multirow{2}{*}{***} \\
 &
   &
  Pocket &
  0.8 &
  0.988 (0.002) &
  0.987 &
   &
  0.000 (0.000) &
  0.000 &
   &
  0.451 (0.133) &
  0.423 &
   \\ \cline{2-13} 
 &
  \multirow{2}{*}{ResNet} &
  Full &
  91.9 &
  0.988 (0.002) &
  0.989 &
  \multirow{2}{*}{***} &
  0.000 (0.000) &
  0.000 &
  \multirow{2}{*}{NS} &
  0.329 (0.090) &
  0.313 &
  \multirow{2}{*}{***} \\
 &
   &
  Pocket &
  0.8 &
  0.988 (0.002) &
  0.989 &
   &
  0.000 (0.000) &
  0.000 &
   &
  0.405 (0.124) &
  0.392 &
   \\ \cline{2-13} 
 &
  \multirow{2}{*}{DenseNet} &
  Full &
  135.9 &
  0.988 (0.002) &
  0.989 &
  \multirow{2}{*}{**} &
  0.000 (0.000) &
  0.000 &
  \multirow{2}{*}{NS} &
  0.441 (0.159) &
  0.401 &
  \multirow{2}{*}{NS} \\
 &
   &
  Pocket &
  1.3 &
  0.988 (0.002) &
  0.989 &
   &
  0.000 (0.000) &
  0.000 &
   &
  0.446 (0.166) &
  0.401 &
   \\ \cline{2-13} 
 &
  \multirow{2}{*}{HRNet} &
  Full &
  6.5 &
  0.988 (0.002) &
  0.989 &
  \multirow{2}{*}{***} &
  0.000 (0.000) &
  0.000 &
  \multirow{2}{*}{NS} &
  0.412 (0.141) &
  0.373 &
  \multirow{2}{*}{NS} \\
 &
   &
  Pocket &
  0.7 &
  0.988 (0.002) &
  0.989 &
   &
  0.000 (0.000) &
  0.000 &
   &
  0.411 (0.137) &
  0.380 &
   \\ \cline{2-13} 
 &
  \multirow{2}{*}{nnUNet} &
  Full &
  31.2 &
  0.989 (0.002) &
  0.990 &
  \multirow{2}{*}{***} &
  0.000 (0.000) &
  0.000 &
  \multirow{2}{*}{NS} &
  0.272 (0.049) &
  0.260 &
  \multirow{2}{*}{***} \\
 &
   &
  Pocket &
  0.8 &
  0.989 (0.002) &
  0.989 &
   &
  0.000 (0.000) &
  0.000 &
   &
  0.284 (0.045) &
  0.276 &
   \\ \hline
\multicolumn{3}{@{}p{1.5in}}{\footnotesize NS, p-value $> 0.05$}\\
\multicolumn{3}{@{}p{1.5in}}{\footnotesize *, p-value $< 0.05$}\\
\multicolumn{3}{@{}p{1.5in}}{\footnotesize **, p-value $< 0.01$}\\  
\multicolumn{3}{@{}p{1.5in}}{\footnotesize ***, p-value $< 0.001$}\\
\end{tabular}%
}
\caption{Accuracy of deep learning models on a set of medical imaging tasks for Pocket vs. full-sized architectures. The PocketNet models' accuracy scores are generally comparable (within 1\% or less) of the full models' accuracy scores. Accuracy scores for segmentation tasks (LiTS, NFBS, BraTS) are evaluated using Dice Similarity Coefficient, Hausdorff 95 distance, and average surface distance.}
\label{fig:accuracy-table-seg}
\end{table*}

\begin{table}[]
\centering
\resizebox{\columnwidth}{!}{%
\begin{tabular}{lllcc}
\hline
\multicolumn{1}{c}{Task} & \multicolumn{1}{c}{Architecture} & \multicolumn{1}{c}{Variant} & \# Parameters (M) & AUC \\ \hline
\multirow{3}{*}{COVIDx} & U-Net Encoder & Full & 1.2 & 0.997 \\
 & U-Net Encoder & Pocket & 0.02 & 0.999 \\
 & State-of-the-art \cite{covidx} & Full & 8.8 & 0.994 \\ \hline
\end{tabular}%
}
\caption{Accuracy of deep learning models on the COVIDx classification task for Pocket vs. full-sized architectures. The PocketNet models' accuracy scores are generally comparable (within 1\% or less) to the full models' accuracy scores. Accuracy scores for the COVIDx classification task are evaluated using AUC, and hence do not have standard deviations or p-values. \label{tab:accuracy-table-class}}
\end{table}

For the BraTS and LiTS datasets, we generally do not see significant performance differences (p $<$ 0.05 [Wilcoxon signed-rank test]) between the full and pocket architectures. In the cases where we see a significant difference in performance, the differences are small, with median Dice scores differing by less than 1\% and median Hausdorff distances differing by less than a fraction of a millimeter. We also do not see a clear pattern in significant performance differences between each pocket and full architecture. Given this and the fact that these differences are generally small, we believe that the outperformance of a full-sized network by its pocket counterpart or vice versa can be explained by the stochastic nature of training each network. 

For the NFBS dataset, we generally see statistically significant differences in performance. However, we again note that these differences are small, and there is no clear pattern for which architecture (i.e., full or pocket) outperforms. These differences are so small (less than 0.0001 for Dice) that they bear no clinical significance. Such minor differences, in this case, are imperceptible and meaningless in a practical setting.

For all three segmentation tasks, these insignificant or minor differences in performance indicate a reduction in the number of parameters by more than an order of magnitude.

\subsection{Performance Analysis}
\label{sec:gpu-usage}
Using the training parameters described in Section \ref{sec:training}, we profile the training performance of a full U-Net and a Pocket U-Net using the BraTS dataset. Namely, we measure peak GPU memory utilization during training and the average time per training step for varying batch sizes for each network using the TensorFlow Profiler \cite{tensorflow}. To ensure accurate comparisons of performance, we conduct these experiments on a Google Colaboratory notebook with a dedicated NVIDIA Tesla T4 GPU with 16 GB of available memory. 

\begin{figure*}[ht!]
\centering
\includegraphics[width=\textwidth]{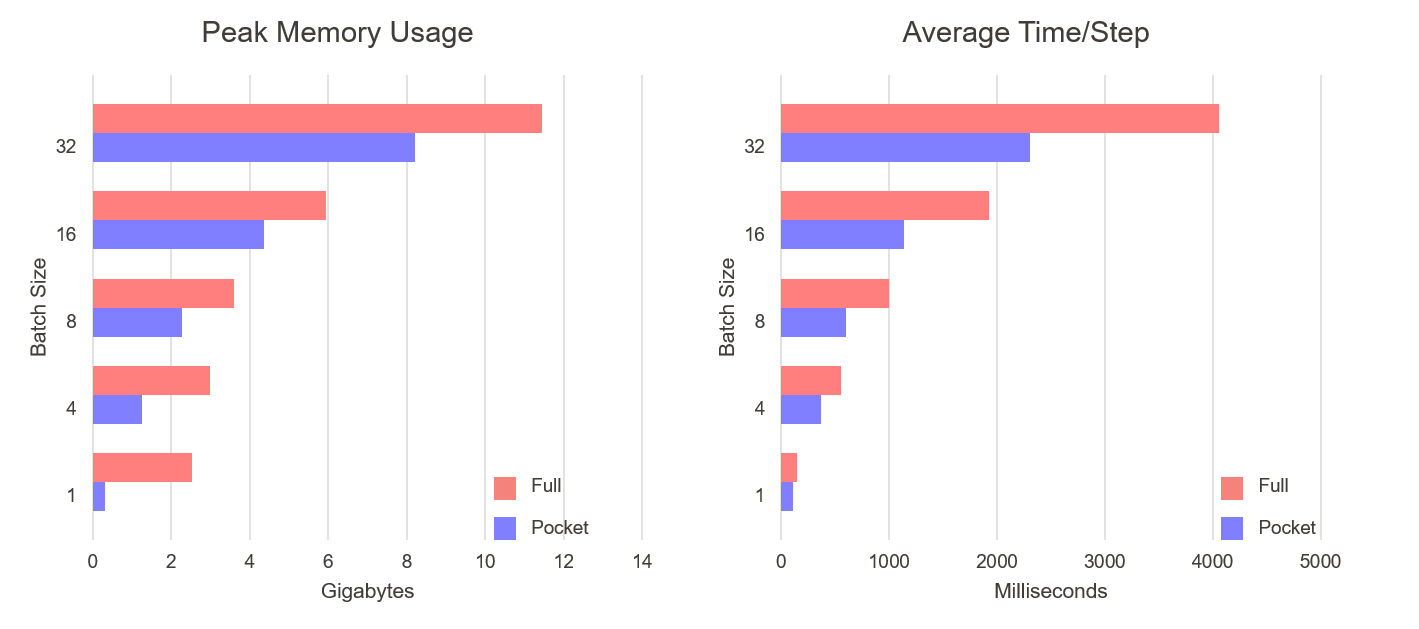}
\caption{(Left) Peak memory usage during training on the BraTS dataset for Pocket U-Net vs Full U-Net for varying batch sizes. The PocketNet architecture results in memory savings of between 28.3\% and 87.7\%. (Right) Average time per training step on BraTS dataset for Pocket U-Net vs Full U-Net for varying batch sizes. The PocketNet models speed up the average time per training step by between 25.0\% and 43.2\%. \label{fig:performance-analysis}}
\end{figure*}

The GPU memory usage and training time per step for this experiment are shown in Figure \ref{fig:performance-analysis}; we see that our PocketNet architecture reduces memory usage and speeds up training time for every batch size. Specifically, the Pocket U-Net reduces the peak memory usage for training by between 28.3\% and 87.7\%, with smaller batch sizes resulting in greater savings. This relationship is possibly due to the increasing portion of GPU memory allocated for storing data as the batch size increases. The PocketNet models improve the average time per training step by between 25.0\% and 43.2\%, with larger batches yielding greater time savings. This behavior may be due to larger batch sizes taking advantage of the computational parallelism of modern GPUs.

In addition to training performance, we benchmark the inference speed of full-sized networks and their PocketNet counterparts. We choose throughput in terms of the number of images predicted per second as our metric for inference speed. Throughput is a popular way of benchmarking the inference speed of a network \cite{throughput1, throughput2, throughput3, throughput4}. Higher throughput means that a network can perform faster inference in a given computing environment. We measure the throughput of the full and pocket variants for several networks and compute the percent speed up of the PocketNet for several different image sizes. The percent speed up is given by the following:
\begin{align*}
    \text{\% speed up } = 100 \times \frac{\text{pocket throughput } -  \text{ full throughput}}{\text{full throughput}}
\end{align*}
Table \ref{tab:infer} shows the inference throughput of each architecture for various image sizes. For architectures with a standard U-Net backbone (i.e., U-Net, ResNet, and DenseNet), we see modest improvements in inference speed for the PocketNet variants that range from 1\% to 12\%. A possible explanation for these minor improvements in inference speed is the highly parallelized computation of convolution operators on modern GPUs. For HRNet, we see more significant improvements in inference throughput ranging from 12\% to 17\%. Within the HRNet architecture, we see more non-convolution operations like upsampling, downsampling, and addition than in U-shaped architectures, which may explain the more significant increases in throughput for the Pocket HRNet architecture. 


\begin{table}[ht!]
\centering
\resizebox{0.45\textwidth}{!}{%
\begin{tabular}{cllcc}
\hline
Image Size & \multicolumn{1}{c}{Architecture} & \multicolumn{1}{c}{Variant} & \begin{tabular}[c]{@{}c@{}}Throughput \\ (image/sec)\end{tabular} & \% Speed Up \\ \hline
\multirow{8}{*}{32$\times$32$\times$32} & \multirow{2}{*}{U-Net} & Full & 17.86 & \multirow{2}{*}{5.89} \\
 &  & Pocket & 18.98 &  \\ \cline{2-5} 
 & \multirow{2}{*}{ResNet} & Full & 16.67 & \multirow{2}{*}{1.17} \\
 &  & Pocket & 16.86 &  \\ \cline{2-5} 
 & \multirow{2}{*}{DenseNet} & Full & 14.37 & \multirow{2}{*}{3.74} \\
 &  & Pocket & 14.93 &  \\ \cline{2-5} 
 & \multirow{2}{*}{HRNet} & Full & 13.16 & \multirow{2}{*}{12.24} \\
 &  & Pocket & 14.99 &  \\ \hline
\multirow{8}{*}{64$\times$64$\times$64} & \multirow{2}{*}{U-Net} & Full & 15.50 & \multirow{2}{*}{2.48} \\
 &  & Pocket & 15.90 &  \\ \cline{2-5} 
 & \multirow{2}{*}{ResNet} & Full & 12.94 & \multirow{2}{*}{12.54} \\
 &  & Pocket & 14.79 &  \\ \cline{2-5} 
 & \multirow{2}{*}{DenseNet} & Full & 12.61 & \multirow{2}{*}{11.60} \\
 &  & Pocket & 14.27 &  \\ \cline{2-5} 
 & \multirow{2}{*}{HRNet} & Full & 10.53 & \multirow{2}{*}{16.53} \\
 &  & Pocket & 12.61 &  \\ \hline
\multirow{8}{*}{128$\times$128$\times$128} & \multirow{2}{*}{U-Net} & Full & 7.84 & \multirow{2}{*}{5.88} \\
 &  & Pocket & 8.33 &  \\ \cline{2-5} 
 & \multirow{2}{*}{ResNet} & Full & 6.72 & \multirow{2}{*}{3.70} \\
 &  & Pocket & 6.98 &  \\ \cline{2-5} 
 & \multirow{2}{*}{DenseNet} & Full & 5.38 & \multirow{2}{*}{7.86} \\
 &  & Pocket & 5.84 &  \\ \cline{2-5} 
 & \multirow{2}{*}{HRNet} & Full & 3.89 & \multirow{2}{*}{13.41} \\
 &  & Pocket & 4.49 &  \\ \hline
\end{tabular}%
}
\caption{Inference throughput for PocketNet vs. full architectures over multiple image sizes using a single NVIDIA RTX 8000 GPU. In this case, throughput measures the number of images predicted per second. Higher throughput implies faster inference speed. In every case, we observe faster inference times with pocket models, with speed-ups (see formula above) varying from 1 to 17\%.}
\label{tab:infer}
\end{table}

\subsection{Ablation Study}
\label{sec:ablation}
To assess the effects of feature map doubling in U-shaped architectures, we perform an ablation study on a standard U-net using the LiTS dataset. In the first iteration of the ablation study, we start with a Full U-Net where we double the number of feature maps at every resolution level. In the next iteration, we construct a U-Net where we stop doubling feature maps after the second-to-last resolution level (i.e., 8 $\times$ 8 $\times$ 8 at $d = 1$). We continue until we arrive at the Pocket U-Net. By performing this ablation study, we can determine at what resolution does feature map doubling become important for the network's accuracy.

For each of these networks, we perform a five-fold cross-validation using training and inference parameters described in Sections \ref{sec:training} and \ref{sec:inference}. Table \ref{tab:ablation} shows the results of each iteration. In every case, we see small differences in the distribution of the resulting Dice scores. This small difference in performance among iterations suggests that doubling the number of feature maps at each resolution level might be unnecessary.

\begin{table*}[ht!]
\centering
\resizebox{0.8\textwidth}{!}{%
\begin{tabular}{lcccc}
\hline
\multicolumn{1}{c}{\multirow{2}{*}{\begin{tabular}[c]{@{}c@{}}Stop feature map doubling \\ after depth $d$\end{tabular}}} &
  \multirow{2}{*}{\begin{tabular}[c]{@{}c@{}}Max features \\ per convolution\end{tabular}} &
  \multirow{2}{*}{\begin{tabular}[c]{@{}c@{}}\# Parameters \\ (M)\end{tabular}} &
  \multicolumn{2}{c}{Dice} \\ \cline{4-5} 
\multicolumn{1}{c}{}   &      &      & Mean $\pm$ Std.   & Median \\ \hline
$d=0$ i.e. Full U-Net  & 1024 & 90.5 & 0.922 $\pm$ 0.098 & 0.952  \\
$d = 1$                & 512  & 60.0 & 0.921 $\pm$ 0.109 & 0.953  \\
$d = 2$                & 256  & 24.3 & 0.923 $\pm$ 0.069 & 0.952  \\
$d = 3$                & 128  & 8.40 & 0.907 $\pm$ 0.145 & 0.950  \\
$d = 5$                & 64   & 2.60 & 0.922 $\pm$ 0.105 & 0.951  \\
$d = 1$ i.e. PocketNet & 32   & 0.80 & 0.930 $\pm$ 0.059 & 0.952  \\ \hline
\end{tabular}%
}
\caption{Accuracy of U-Net architectures on LiTS dataset where feature map doubling stops after a given depth $d$. In every case, we see small differences in the distribution of the resulting Dice scores. \label{tab:ablation}}
\end{table*}


\subsection{Feature Activation Analysis}
\label{sec:feature-activation}
\begin{figure*}[ht!]
\centering
\includegraphics[width=\textwidth]{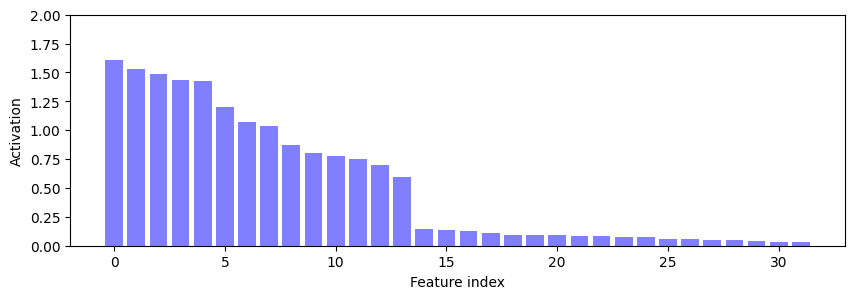}
\\
\includegraphics[width=\textwidth]{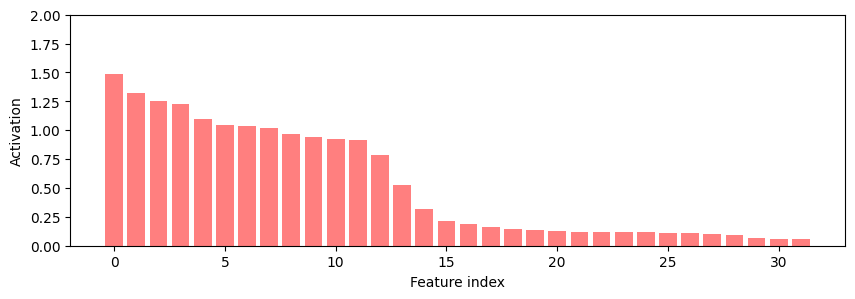}
\caption{Feature map activations in final layer before voxel-wise classification in a Pocket U-Net (top) and full U-Net (bottom) for the BraTS dataset. In both cases, we see a similar number of features being activated with roughly the same intensities.   \label{fig:activations}}
\end{figure*}

We conjecture that the comparable performance between the PocketNet and full architectures is due to both networks having similar representation capabilities, that ultimately both networks build similar representations of the image data as they compute the final segmentations. To test whether the networks learn similar features, we look at the mean of the feature map activations in the final layer before voxel-wise classification in full and Pocket U-Nets trained on BraTS data. In this way, we can determine if both networks are capturing similar features and information. For each image in the test set, we examine the output of the activation functions from the final layer before classification and measure the size of the response towards the convolution feature, by taking the channel-wise average of the resulting feature maps of the activation output. This process is repeated for the entire test set. More precisely, let $F_i$ be the average of the $i^{th}$ feature map resulting from activation outputs from the final layer before classification over the entire test set. Let $f_i^j$ be the $i^{th}$ feature map resulting from the activation functions for test patch $j$. For each $i = 1, \dots, 32$, we have
\begin{align*}
    F_i = \frac{1}{N_{test}} \sum_{j = 1}^{N_{test}} \frac{1}{V^j} \sum_{k=1}^{V^j} \left( f_i^j \right)_k,
\end{align*}
where $N_{test}$ is the number patches in the test set, $V^j$ is the volume of test patch $j$, and $\left( f_i^j \right)_k$ is the intensity of the $k^{th}$ voxel in $f_i^j$. Figure \ref{fig:activations} shows the averages of the resulting feature maps. We see a similar number of features being activated with similar intensities for both cases. This similarity suggests that the full and Pocket U-Nets learn similar latent feature representations used for the final voxel-wise classification. Note that we sort the mean feature activations from highest to lowest for visual purposes. This order does not matter because the indexing in any hidden layer can always be permuted by the next layer. 

\subsection{Model Saturation}
\label{sec:saturation}
\begin{figure*}[ht!]
\centering
\begin{subfigure}[t]{0.48\textwidth}
    \centering
    \includegraphics[width=\textwidth]{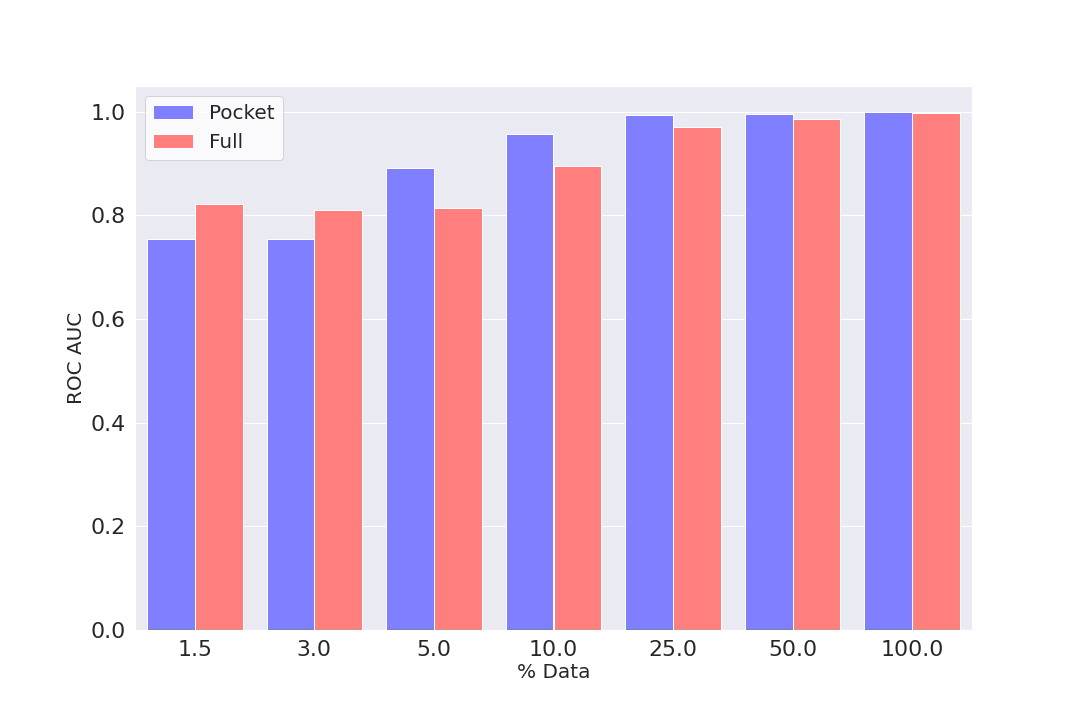}
    \subcaption{Classification performance on COVIDx8B data for PocketNet vs. Standard U-Net encoder classifier, trained using various subsets of the training data. 
    \label{fig:data-saturation-covid}}
\end{subfigure}~
\begin{subfigure}[t]{0.48\textwidth}
    \centering
    \includegraphics[width=\textwidth]{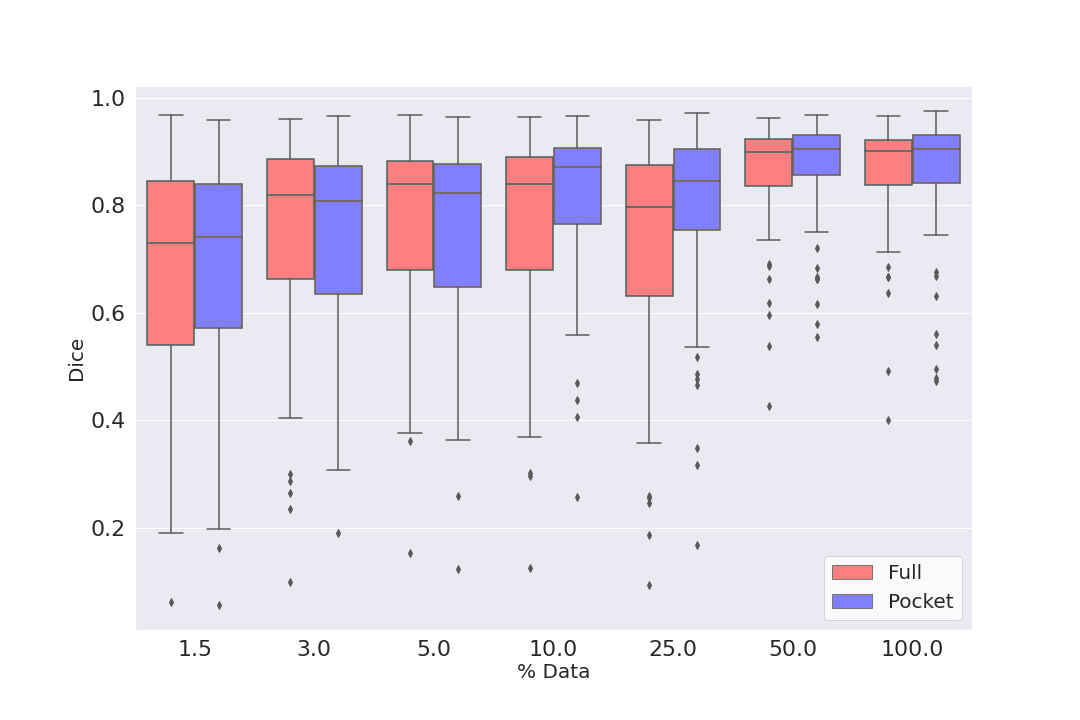}
    \subcaption{Distributions of Dice scores on BraTS test set for a full U-Net and a pocket U-Net for varying training set sizes.
    \label{fig:brats-violins}}
\end{subfigure}
\begin{subfigure}[t]{0.48\textwidth}
    \centering
    \includegraphics[width=\textwidth]{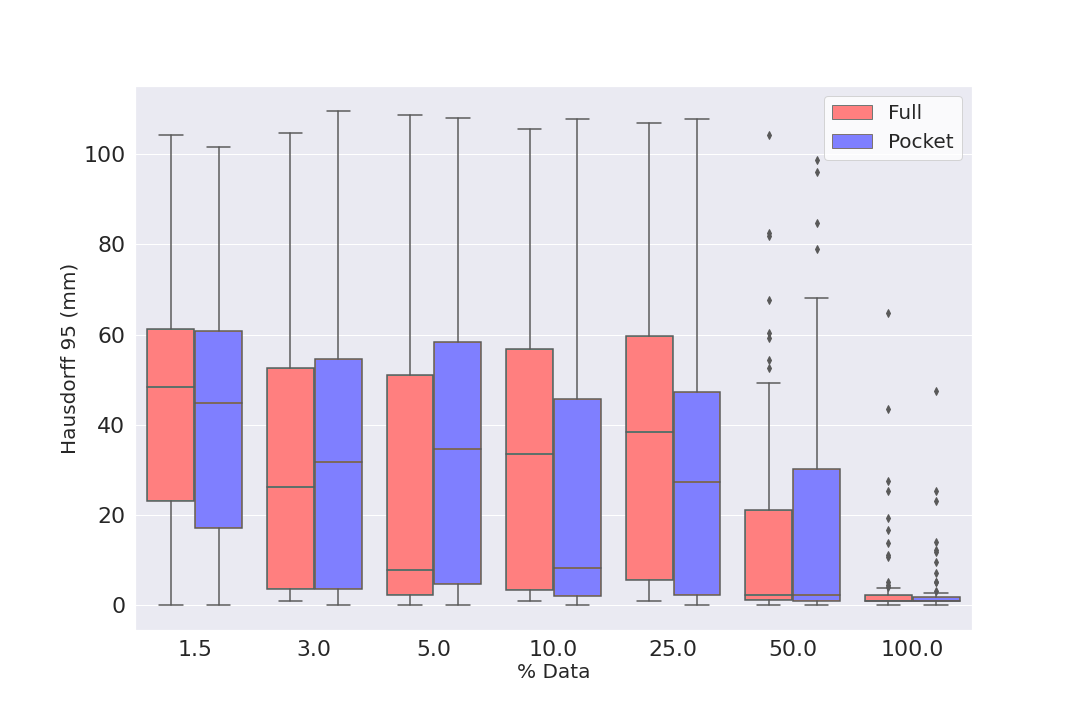}
    \subcaption{Distributions of Hausdorff 95 distances on BraTS test set for a full U-Net and a pocket U-Net for varying training set sizes.
    \label{fig:brats-sat-haus}}
\end{subfigure}
\begin{subfigure}[t]{0.48\textwidth}
    \centering
    \includegraphics[width=\textwidth]{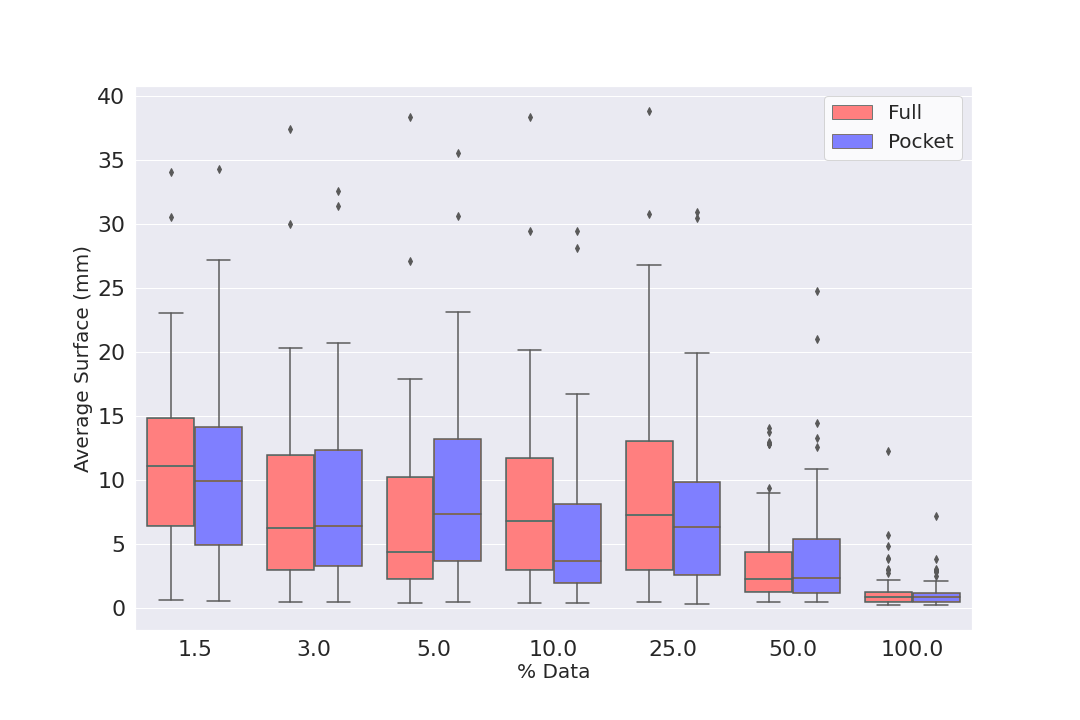}
    \subcaption{Distributions of average surface distances on BraTS test set for a full U-Net and a pocket U-Net for varying training set sizes.
    \label{fig:brats-sat-avg-surf}}
\end{subfigure}
\caption{Testing results from using small subsets of data for PocketNet vs. full architectures on COVID19x8B classification and BraTS segmentation challenges. \label{data-saturation}}
\end{figure*}

A possible concern is that PocketNet models, due to their reduced parameter count, could saturate earlier during training than do full-sized architectures, which could result in the comparable performance we observe in our results in Section \ref{sec:acc}. To test this, we repeat the experiments described above for the COVIDx8B and BraTS data challenges using successively less data in the training set. For every iteration, we keep a fixed validation and test set. For the COVIDx8B dataset, we fix 10\% of the training data as a validation set and use the original test set. Similarly, for the BraTS data, we take 20\% of the training data as a test set and use 10\% of the remaining patients as a validation set. Additionally, we do not use data augmentation for this particular experiment. The results of this are shown in Figure \ref{fig:data-saturation-covid} and Figure \ref{fig:brats-violins}.

In Figure \ref{fig:data-saturation-covid}, we see that the AUC values increase for both the PocketNet and full architectures as the size of the training set increases. Furthermore, we observe that these AUC values plateau at 1.0 (i.e., perfect prediction), and the PocketNet classifier saturates sooner than its full-sized counterpart. These observations suggest that the reduced architecture resulting from the PocketNet paradigm learns faster with fewer data points than its full-sized counterpart. Similarly, Figures \ref{fig:brats-violins}, \ref{fig:brats-sat-haus}, and \ref{fig:brats-sat-avg-surf} show that the segmentation accuracy of the full and pocket U-Net architectures improves as the training set size increases. Both architecture types show the expected improvement in performance with each increase in the dataset size, plateauing to similar distributions.

\section{Discussion}
\label{sec:discussion}

Our results show that large numbers of parameters (millions or tens of millions) may not be necessary for deep learning in medical image analysis, as comparable performance is achievable with substantially smaller networks using the same architectures but without doubling the number of channels at coarser resolutions. 
This suggests that overparameterization, which is increasingly regarded as a key reason why neural networks learn efficiently, might not be as critical as previously suggested \cite{allen2019convergence, arora2018optimization, rice2020overfitting}. However, we note that our PocketNets may still be are overparameterized, and the combination of our proposed PocketNet paradigm with other model reduction techniques should be explored. For example, replacing the traditional convolution layers with DS convolution layers in our Pocket ResNet for LiTS liver segmentation further reduces the number of parameters to roughly 10,000. Pruning an already trained PocketNet model may also potentially yield further parameter reductions. 

\begin{figure}[ht!]
    \centering
    \includegraphics[width=\columnwidth]{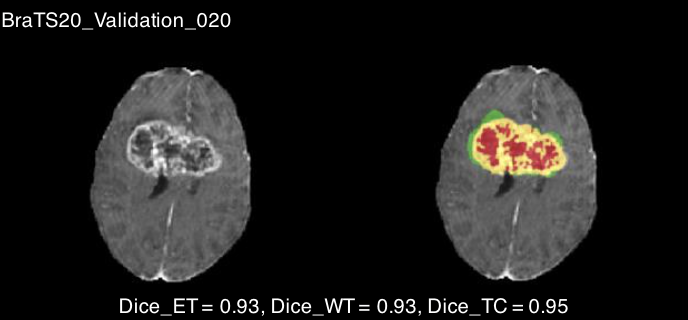}
    \caption{Multi-class Pocket DenseNet segmentation on BraTS 2020 Validation image. Enhancing Tumor (ET), Whole Tumor (WT), and Tumor Core (TC) Dice scores are in line with state-of-the-art predictions.\label{fig:brats-multi}}
\end{figure}

The deep learning tasks presented in this study are all single-label segmentation or binary classification. The goal of ongoing and future work using the PocketNet paradigm is to test this approach on more complex domains such as BraTS multi-class tumor segmentation and LiTS tumor segmentation. Figure \ref{fig:brats-multi} shows an example of a multi-class segmentation prediction mask produced by a Pocket DenseNet. Our results for PocketNet architectures applied to the BraTS multi-class segmentation task are available at \url{https://www.cbica.upenn.edu/BraTS20/lboardValidation.html} under the team name ``aecmda'' and will be updated periodically. 

\begin{figure*}[ht!]
    \centering
    \includegraphics[width=\textwidth]{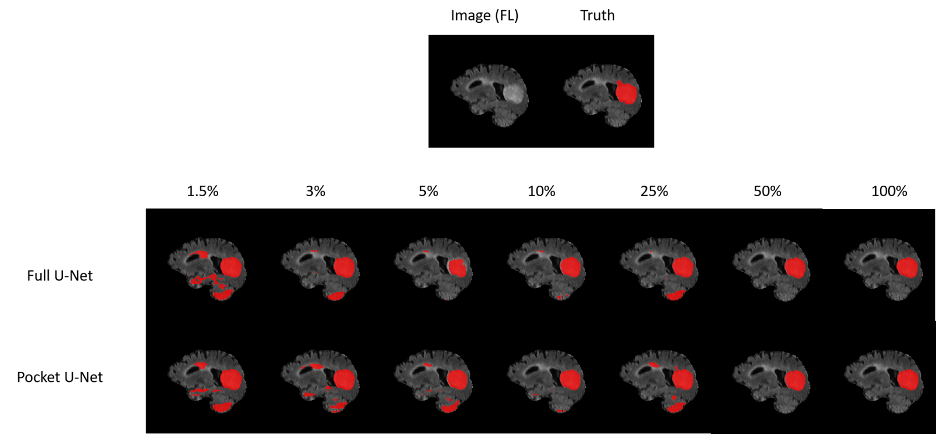}
    \caption{Example predictions for full and pocket U-Net on BraTS dataset with varying percentages of the dataset used for training. In both cases, we see similar behavior, with both architectures producing noisy predictions with fewer data and more accurate predictions as we increase the dataset size. \label{example-sat}}
\end{figure*}

Regarding Section \ref{sec:saturation} and the results shown in Figure \ref{data-saturation}, we note that the results for each task using 1.5\% of the data look surprisingly good. For the segmentation task (i.e., BraTS tumor segmentation), a closer analysis reveals that while the full and pocket U-Nets can identify brain tumors in the BraTS dataset with a small percentage of the available training data, they tend to produce rather noisy predictions. Figure \ref{example-sat} illustrates an example of these predictions. In both cases, the networks erroneously predict brighter areas of the MR images as tumors. We hypothesize that the images in the BraTS dataset are similar enough that even with a small percentage of available data, each network can learn to predict rough segmentations of the target. This behavior, however, supports our claim that the pocket and full-sized U-Nets behave similarly with varying dataset sizes. Again, Figure \ref{example-sat} illustrates this by showing an example prediction from each network for small and large training sets. In both cases, we see similar behavior, with both architectures producing noisy predictions with fewer data and more accurate predictions as we increase the dataset size.

Additionally, the Dice coefficient may be too forgiving of a metric for this type of analysis. Even for noisy predictions, we can achieve Dice scores of about 0.7. To mitigate this, we compute the Hausdorff 95 and average surface distances for each set of predictions in Figures \ref{fig:brats-sat-haus} and \ref{fig:brats-sat-avg-surf}, respectively. With the Hausdorff 95 metric, we see that predictions from training on small datasets generally result in significant errors with substantial variation. However, for larger datasets, we see that for each network variant, the distribution of Hausdorff 95 distances converges to smaller values with reduced variance. We again see similar behavior with the average surface distance. We see significant errors with considerable variance for each network variant for smaller datasets. These distributions plateau to smaller values with less variance with increased training set size for both full and pocket networks. These cases support our initial claim that both networks exhibit similar behavior with different dataset sizes.  

For the classification results in Figure \ref{data-saturation}, we again argue that the images in the dataset are similar enough to each other that even with a small percentage taken as a training set, each architecture can roughly discriminate between the two classes.

When we employ PocketNet models, we achieve similar performance to full-sized networks while enjoying the advantages of faster training times and lower memory requirements. The smaller models produced by our proposed PocketNet paradigm can potentially lower the entry costs (computational and monetary) of training deep learning models in resource-constrained environments without access to specialized computing equipment. With less GPU memory required for training, cheaper hardware can be purchased, or less expensive cloud computing instances can be used to train deep learning models for medical image analysis. The faster training times for PocketNets can also reduce costs by reducing the number of hours spent training models on cloud computing instances.


As to \emph{why} the PocketNet models perform at least comparable to their counterpart full models, the similarity in intensities of each signal in the final layer activation maps suggests that the models ultimately learn the same representations. Despite the reduced number of parameters, the approximation space that the Pocket architectures can represent is comparable to the approximation space that the full models can achieve. This conjecture is supported by the ablation study results in Table \ref{tab:ablation}: the median Dice scores for e.g. U-Net is nearly identical, regardless of the depth that the feature map stops doubling. In this study, the additional features supplied at depths where the doubling continued did not improve the model's performance, as the approximation spaces are all similar regardless of whether the number of features was doubled at that depth. This ablation study suggests that, since there is no increased benefit of having larger models with the number of features doubling per layer, that for these medical imaging problems, doubling the number of channels is unnecessary and PocketNet models can be used to achieve comparable accuracy instead. Since these smaller models are just as expressive (and capable) as their full counterparts, these models can be trained (and later, deployed) with cheaper hardware or by provisioning smaller cloud instances, saving time, money, and effort by institutions performing deep learning medical image analysis.

\bibliographystyle{ieeetr}
\bibliography{sources}

\begin{thebibliography}{10}

\bibitem{intro-2}
N.~Sharma and et~al., ``Automated medical image segmentation techniques,'' {\em
  Journal of Medical Physics}, vol.~35, pp.~3--14, 1 2010.

\bibitem{intro-1}
D.~Thompson and et~al., ``Evaluation of an automatic segmentation algorithm for
  definition of head and neck organs at risk,'' {\em Radiation Oncology},
  vol.~9, pp.~1--12, 8 2014.

\bibitem{intro-3}
E.~Ermis and et~al., ``Fully automated brain resection cavity delineation for
  radiation target volume definition in glioblastoma patients using deep
  learning,'' {\em Radiation Oncology}, vol.~15, pp.~1--10, 5 2020.

\bibitem{unet}
O.~Ronneberger, P.~Fischer, and T.~Brox, ``U-net: Convolutional networks for
  biomedical image segmentation,'' in {\em International Conference on Medical
  image computing and computer-assisted intervention}, pp.~234--241, Springer,
  2015.

\bibitem{unet-3d}
{\"O}.~{\c{C}}i{\c{c}}ek, A.~Abdulkadir, S.~S. Lienkamp, T.~Brox, and
  O.~Ronneberger, ``3d u-net: learning dense volumetric segmentation from
  sparse annotation,'' in {\em International conference on medical image
  computing and computer-assisted intervention}, pp.~424--432, Springer, 2016.

\bibitem{densenet}
G.~Huang, Z.~Liu, L.~Van Der~Maaten, and K.~Q. Weinberger, ``Densely connected
  convolutional networks,'' in {\em Proceedings of the IEEE conference on
  computer vision and pattern recognition}, pp.~4700--4708, 2017.

\bibitem{resnet}
K.~He, X.~Zhang, S.~Ren, and J.~Sun, ``Identity mappings in deep residual
  networks,'' in {\em European conference on computer vision}, pp.~630--645,
  Springer, 2016.

\bibitem{lambda}
``Gpu cloud, workstations, servers, laptops for deep learning.''
  \url{https://lambdalabs.com/}.

\bibitem{aws}
``Amazon aws ec2 pricing.'' \url{https://aws.amazon.com/ec2/pricing/}.

\bibitem{lenet}
Y.~Lecun, L.~Bottou, Y.~Bengio, and P.~Haffner, ``Gradient-based learning
  applied to document recognition,'' {\em Proceedings of the IEEE}, vol.~86,
  no.~11, pp.~2278--2324, 1998.

\bibitem{imagenet}
A.~Krizhevsky, I.~Sutskever, and G.~E. Hinton, ``Imagenet classification with
  deep convolutional neural networks,'' {\em Advances in neural information
  processing systems}, vol.~25, 2012.

\bibitem{alexnet}
A.~Krizhevsky, I.~Sutskever, and G.~E. Hinton, ``Imagenet classification with
  deep convolutional neural networks,'' {\em Commun. ACM}, vol.~60, p.~84–90,
  may 2017.

\bibitem{hrnet}
K.~Sun, B.~Xiao, D.~Liu, and J.~Wang, ``Deep high-resolution representation
  learning for human pose estimation,'' in {\em CVPR}, 2019.

\bibitem{nnunet}
F.~Isensee, P.~F. Jaeger, S.~A. Kohl, J.~Petersen, and K.~H. Maier-Hein,
  ``nnu-net: a self-configuring method for deep learning-based biomedical image
  segmentation,'' {\em Nature Methods 2020 18:2}, vol.~18, pp.~203--211, 12
  2020.

\bibitem{unet-common}
N.~Siddique, S.~Paheding, C.~P. Elkin, and V.~Devabhaktuni, ``U-net and its
  variants for medical image segmentation: A review of theory and
  applications,'' {\em IEEE Access}, vol.~9, pp.~82031--82057, 2021.

\bibitem{pruning1}
Y.~Lecun, J.~Denker, and S.~Solla, ``Optimal brain damage,'' {\em Advances in
  Neural Information Processing Systems}, vol.~2, pp.~598--605, 01 1989.

\bibitem{ds-conv-primer}
R.~Ye, F.~Liu, and L.~Zhang, ``3d depthwise convolution: Reducing model
  parameters in 3d vision tasks,'' in {\em Advances in Artificial
  Intelligence}, pp.~186--199, Springer International Publishing, 2019.

\bibitem{filter-reduce1}
J.~van~der Putten, F.~van~der Sommen, and P.~H.~N. de~With, ``Influence of
  decoder size for binary segmentation tasks in medical imaging,'' in {\em
  Medical Imaging 2020: Image Processing}, vol.~11313, pp.~276 -- 281,
  International Society for Optics and Photonics, SPIE, 2020.

\bibitem{unet++-pruning}
Z.~Zhou and et~al., ``Unet++: A nested u-net architecture for medical image
  segmentation,'' in {\em Deep Learning in Medical Image Analysis and
  Multimodal Learning for Clinical Decision Support}, pp.~3--11, Springer
  International Publishing, 2018.

\bibitem{NASUnet-pruning}
Y.~Weng, T.~Zhou, Y.~Li, and X.~Qiu, ``Nas-unet: Neural architecture search for
  medical image segmentation,'' {\em IEEE Access}, vol.~7, pp.~44247--44257,
  2019.

\bibitem{medseg-pruning}
A.~Feng-Ping and L.~Zhi-Wen, ``Medical image segmentation algorithm based on
  feedback mechanism convolutional neural network,'' {\em Biomedical Signal
  Processing and Control}, vol.~53, p.~101589, 2019.

\bibitem{xception}
F.~Chollet, ``Xception: Deep learning with depthwise separable convolutions,''
  in {\em Proceedings of the IEEE conference on computer vision and pattern
  recognition}, pp.~1251--1258, 2017.

\bibitem{mobilenet}
A.~G. Howard and et~al., ``Mobilenets: Efficient convolutional neural networks
  for mobile vision applications,'' {\em CoRR}, vol.~abs/1704.04861, 2017.

\bibitem{ds-conv-1}
N.~Alalwan, A.~Abozeid, A.~A. ElHabshy, and A.~Alzahrani, ``Efficient 3d deep
  learning model for medical image semantic segmentation,'' {\em Alexandria
  Engineering Journal}, vol.~60, Issue 1, pp.~1231--1239, 2021.

\bibitem{ds-conv-2}
K.~Qi, H.~Yang, C.~Li, and et~al., ``X-net: Brain stroke lesion segmentation
  based on depthwise separable convolution and long-range dependencies,'' in
  {\em Medical Image Computing and Computer Assisted Intervention -- MICCAI
  2019}, (Cham), pp.~247--255, Springer International Publishing, 2019.

\bibitem{siambook}
T.~F. Chan and J.~J. Shen, {\em Image processing and analysis: variational,
  PDE, wavelet, and stochastic methods}, vol.~94.
\newblock Siam, 2005.

\bibitem{multigrid}
J.~H. Bramble, {\em Multigrid methods}.
\newblock Routledge, 2018.

\bibitem{saad2003iterative}
Y.~Saad, {\em Iterative methods for sparse linear systems}.
\newblock SIAM, 2003.

\bibitem{mgnet}
J.~He and J.~Xu, ``{MgNet: A unified framework of multigrid and convolutional
  neural network},'' {\em Science China Mathematics}, vol.~62, no.~7,
  pp.~1331--1354, 2019.

\bibitem{lits}
P.~Bilic and et~al., ``{The liver tumor segmentation benchmark (LiTS)},'' {\em
  arXiv preprint arXiv:1901.04056}, 2019.

\bibitem{nfbs}
B.~Puccio, J.~P. Pooley, J.~S. Pellman, E.~C. Taverna, and R.~C. Craddock,
  ``{The preprocessed connectomes project repository of manually corrected
  skull-stripped T1-weighted anatomical MRI data},'' {\em GigaScience}, vol.~5,
  p.~45, 10 2016.

\bibitem{brats1}
B.~Menze and et~al., ``The multimodal brain tumor image segmentation benchmark
  (brats),'' {\em IEEE Transactions on Medical Imaging}, vol.~34, no.~10,
  pp.~1993--2024, 2015.

\bibitem{brats2}
S.~Bakas and et~al., ``Identifying the best machine learning algorithms for
  brain tumor segmentation, progression assessment, and overall survival
  prediction in the brats challenge,'' {\em arXiv preprint arXiv:1811.02629},
  2018.

\bibitem{brats3}
S.~Bakas and et~al., ``Advancing the cancer genome atlas glioma mri collections
  with expert segmentation labels and radiomic features,'' {\em Scientific
  Data}, vol.~4, no.~1, p.~170117, 2017.

\bibitem{covidx}
L.~Wang, Z.~Q. Lin, and A.~Wong, ``Covid-net: a tailored deep convolutional
  neural network design for detection of covid-19 cases from chest x-ray
  images,'' {\em Scientific Reports}, vol.~10, p.~19549, Nov 2020.

\bibitem{adam}
D.~P. Kingma and J.~Ba, ``Adam: A method for stochastic optimization,'' {\em
  arXiv preprint arXiv:1412.6980}, 2014.

\bibitem{simpleitk1}
B.~Lowekamp, D.~Chen, L.~Ibanez, and D.~Blezek, ``The design of simpleitk,''
  {\em Frontiers in Neuroinformatics}, vol.~7, p.~45, 2013.

\bibitem{simpleitk2}
Z.~Yaniv, B.~C. Lowekamp, H.~J. Johnson, and R.~Beare, ``Simpleitk
  image-analysis notebooks: a collaborative environment for education and
  reproducible research,'' {\em Journal of Digital Imaging}, vol.~31,
  pp.~290--303, 6 2018.

\bibitem{simpleitk3}
R.~Beare, B.~Lowekamp, and Z.~Yaniv, ``Image segmentation, registration and
  characterization in r with simpleitk,'' {\em Journal of Statistical Software,
  Articles}, vol.~86, no.~8, pp.~1--35, 2018.

\bibitem{actor2020identification}
J.~A. Actor, D.~T. Fuentes, and B.~Rivi{\`e}re, ``Identification of kernels in
  a convolutional neural network: connections between the level set equation
  and deep learning for image segmentation,'' in {\em Medical Imaging 2020:
  Image Processing}, vol.~11313, p.~1131317, International Society for Optics
  and Photonics, 2020.

\bibitem{keras}
F.~Chollet and et~al., ``Keras,'' 2015.
\newblock Software available from keras.org.

\bibitem{tensorflow}
M.~Abadi and et. al., ``{TensorFlow}: Large-scale machine learning on
  heterogeneous systems,'' 2015.
\newblock Software available from tensorflow.org.

\bibitem{throughput1}
K.~Rungsuptaweekoon, V.~Visoottiviseth, and R.~Takano, ``Evaluating the power
  efficiency of deep learning inference on embedded gpu systems,'' in {\em 2017
  2nd International Conference on Information Technology (INCIT)}, pp.~1--5,
  2017.

\bibitem{throughput2}
R.~Xu, F.~Han, and Q.~Ta, ``Deep learning at scale on nvidia v100
  accelerators,'' in {\em 2018 IEEE/ACM Performance Modeling, Benchmarking and
  Simulation of High Performance Computer Systems (PMBS)}, pp.~23--32, 2018.

\bibitem{throughput3}
S.~Mittal, ``A survey on optimized implementation of deep learning models on
  the nvidia jetson platform,'' {\em Journal of Systems Architecture}, vol.~97,
  pp.~428--442, 2019.

\bibitem{throughput4}
L.~Mai, A.~Koliousis, G.~Li, A.-O. Brabete, and P.~Pietzuch, ``Taming
  hyper-parameters in deep learning systems,'' {\em ACM SIGOPS Operating
  Systems Review}, vol.~53, pp.~52--58, 07 2019.

\bibitem{allen2019convergence}
Z.~Allen-Zhu, Y.~Li, and Z.~Song, ``A convergence theory for deep learning via
  over-parameterization,'' in {\em International Conference on Machine
  Learning}, pp.~242--252, PMLR, 2019.

\bibitem{arora2018optimization}
S.~Arora, N.~Cohen, and E.~Hazan, ``On the optimization of deep networks:
  Implicit acceleration by overparameterization,'' in {\em International
  Conference on Machine Learning}, pp.~244--253, PMLR, 2018.

\bibitem{rice2020overfitting}
L.~Rice, E.~Wong, and Z.~Kolter, ``Overfitting in adversarially robust deep
  learning,'' in {\em International Conference on Machine Learning},
  pp.~8093--8104, PMLR, 2020.

\end{thebibliography}

\end{document}